\documentclass[preprint1]{aastex}
\usepackage{amsmath}
\slugcomment{}

\newcommand{\be}{\begin{equation}}
\newcommand{\ee}{\end{equation}}

\shorttitle{The Supernova Hubble Diagram}
\shortauthors{Wei, Wu, Melia \& Maier}

\begin{document}

\title{A Comparative Analysis of the Supernova Legacy Survey\\
Sample with $\Lambda$CDM and the $R_{\rm h}=ct$ Universe$^7$}
\author{Jun-Jie Wei\altaffilmark{1,3}, Xue-Feng Wu\altaffilmark{1,4,5},
Fulvio Melia\altaffilmark{1,2} and Robert S. Maier\altaffilmark{6}}
\altaffiltext{1}{Purple Mountain Observatory, Chinese Academy of Sciences, Nanjing 210008, China.}
\altaffiltext{2}{Department of Physics, The Applied Mathematics Program, and Department of Astronomy,
The University of Arizona, AZ 85721, USA; fmelia@email.arizona.edu.}
\altaffiltext{3}{University of Chinese Academy of Sciences, Beijing 100049, China; jjwei@pmo.ac.cn.}
\altaffiltext{4}{Chinese Center for Antarctic Astronomy, Nanjing 210008, China; xfwu@pmo.ac.cn.}
\altaffiltext{5}{Joint Center for Particle, Nuclear Physics and Cosmology, Nanjing
University--Purple Mountain Observatory, Nanjing 210008, China.}
\altaffiltext{6}{Department of Mathematics, The Statistics Program, and Department of Physics,
The University of Arizona, AZ 85721, USA; rsm@math.arizona.edu}
\altaffiltext{7}{This work is dedicated to the memory of Prof. Tan Lu, who sadly passed away
December 3, 2014. Among his many achievements, he is considered to be one of the founders of
high-energy astrophysics, and a pioneer in modern cosmology, in China.}

\begin{abstract}
The use of Type~Ia SNe has thus far produced the most reliable measurement
of the expansion history of the Universe, suggesting that $\Lambda$CDM
offers the best explanation for the redshift--luminosity distribution
observed in these events.  But the analysis of other kinds of source, such
as cosmic chronometers, gamma ray bursts, and high-$z$ quasars, conflicts
with this conclusion, indicating instead that the constant expansion rate
implied by the $R_{\rm h}=ct$ Universe is a better fit to the data.  The
central difficulty with the use of Type~Ia SNe as standard candles is that
one must optimize three or four nuisance parameters characterizing
supernova luminosities simultaneously with the parameters of an expansion
model.  Hence in comparing competing models, one must reduce the data
independently for each.  We carry~out such a comparison of $\Lambda$CDM and
the $R_{\rm h}=ct$ Universe, using the Supernova Legacy Survey (SNLS)
sample of 252 SN~events, and show that each model fits its individually
reduced data very well.  But since $R_{\rm h}=ct$ has only one free
parameter (the Hubble constant), it follows from a standard model selection
technique that it is to be preferred over $\Lambda$CDM, the minimalist
version of which has three (the Hubble constant, the scaled matter density
and either the spatial curvature constant or the dark-energy
equation-of-state parameter).  We estimate by the Bayes Information
Criterion that in a pairwise comparison, the likelihood of $R_{\rm h}=ct$
is $\sim 90\%$, compared with only $\sim 10\%$ for a minimalist form of
$\Lambda$CDM, in which dark energy is simply a cosmological constant.
Compared to $R_{\rm h}=ct$, versions of the standard model with more
elaborate parametrizations of dark energy are judged to be even less
likely.
\end{abstract}

\section{Introduction}
Type~Ia supernovae have a well-defined luminosity, permitting their use as
standard candles under the assumption that the luminosities of nearby and
distant sources are similarly related to color and light-curve shape.  The
use of such events has thus~far produced the most reliable measurement of
the expansion history of the Universe (Perlmutter et~al.\ 1998, 1999;
Garnavich et~al.\ 1998; Schmidt et~al.\ 1998; Riess et~al.\ 1998), leading
to the discovery of dark energy.

In recent years, several samples of Type~Ia supernova events have been
assembled, and several of these have been merged to form even larger
compilations, such as the Union2.1 sample (Kowalski et~al.\ 2008; Suzuki
et~al.\ 2012), which currently contains 580 supernova detections, including
events at redshift $z>1$ observed with the {\it Hubble} Space Telescope
(see, e.g., Kuznetsova et~al.\ 2008; Dawson et~al.\ 2009; Riess
et~al.\ 2011).  However, while such merged samples offer several
advantages, the fact that each sample has its own set of systematic and
intrinsic uncertainties makes it difficult to fit a cosmological model to
any merged sample.  As we shall discuss, and as Kim (2011) and others have
already pointed~out, it is questionable whether the model parameters can be
estimated by simply minimizing a~$\chi^2$, since parameters characterizing
error dispersion(s) must be estimated simultaneously.  One commonly used
method estimates error dispersion(s) by constraining the reduced~$\chi^2$,
i.e., the $\chi^2$ per degree of freedom, to equal unity. We shall find
that the method of maximum likelihood estimation (MLE) is to be preferred;
but in any method, the presence of multiple dispersion parameters is a
complication.

Fortunately, some of the homogeneous samples are themselves quite large,
and perfectly suited to the type of comparative analysis we wish to
carry~out in this paper.  In~fact, almost half of all the Type~Ia SNe in
the Union2.1 compilation were derived from the single, homogeneous sample
assembled during the first three years of the Supernova Legacy Survey
(SNLS; Guy et~al.\ 2010).  This catalog contains 252 high redshift Type~Ia
supernovae ($0.15<z<1.1$).  The multi-color light-curves of these SNe were
measured using the MegaPrime/MegaCam instrument at the
Canada--France--Hawaii Telescope (CFHT), using repeated imaging of four
one-square degree fields in four bands.  The VLT, Gemini and Keck
telescopes were used to confirm the nature of the SNe and to measure their
redshifts.  Very importantly, since the same instruments and reduction
techniques were employed for all 252 events, it is appropriate to include a
\emph{single} intrinsic dispersion in the analysis of the Hubble Diagram
(HD) constructed from this sample.  The study of catalogs such as this has
led to a general consensus that $\Lambda$CDM offers the best explanation
for the redshift--luminosity relationship, and observational work is now
focused primarily on refining the fits to improve the precision with which
the model parameters are determined.  This is one of the principal
motivations for attempting to merge samples to create catalogs with broader
redshift coverage and better statistics.

But as successful as this program has been, several drawbacks associated
with the use of Type~Ia supernovae have made it necessary to seek
alternative methods of probing the cosmic spacetime.  It is quite difficult
to use supernova measurements in unbiased, comparative studies of competing
expansion histories, since at~least three or four `nuisance' parameters
characterizing the standard candle must be optimized simultaneously with
each model's free parameters, rendering the data compliant to the
underlying theory (Kowalski et~al.\ 2008; Suzuki et~al.\ 2012; Melia
2012a). Several notable attempts have been made to mitigate
the impact of this model dependence, e.g., through the use of
kinematic variables and geometric probes that avoid parametrizing
the fits in terms of pre-assumed model components (Shafieloo
et al. 2012). In the end, however, even these models require
the availability of measurements based on standard candles.
Unfortunately, Type Ia SNe may be used for this purpose only
as long as the nuisance parameters characterizing their
lightcurves are known. The application of such methods to the
Union2.1 Type Ia SN sample uses nuisance parameters optimized
for the concordance model, so the current results are not
completely free of any biases.

The expansion of the Universe is now being studied by several other
methods, including the use of cosmic chronometers (Jimenez \& Loeb 2002;
Simon et~al.\ 2005; Stern et~al.\ 2010; Moresco et~al.\ 2012; Melia \&
Maier 2013), gamma ray bursts (GRBs; Norris et~al.\ 2000; Amati
et~al.\ 2002; Schaefer 2003; Wei \& Gao 2003; Yonetoku et~al.\ 2004;
Ghirlanda et~al.\ 2004; Liang \& Zhang 2005; Liang et~al.\ 2008; Wang
et~al.\ 2011; Wei et~al.\ 2013), and high-$z$ quasars (Kauffmann \&
Haehnelt 2000; Wyithe \& Loeb 2003; Hopkins et~al.\ 2005; Croton
et~al.\ 2006; Fan 2006; Willott et~al.\ 2003; Jiang et~al.\ 2007; Kurk
et~al.\ 2007; Tanaka \& Haiman 2009; Lippai et~al.\ 2009; Hirschmann
et~al.\ 2010; Melia 2013a).  In contrast to the perception based on Type~Ia
SNe that $\Lambda$CDM can best account for the observed expansion of the
Universe, the conclusion from these other studies is that the cosmic
dynamics is better described by a cosmology we refer to as the $R_{\rm
  h}=ct$ Universe (Melia 2007; Melia \& Abdelqader 2009; Melia \& Shevchuk
2012).

For example, a comparative analysis was recently carried out of the
$\Lambda$CDM and $R_{\rm h}=ct$ cosmologies using the GRB Hubble diagram
(Wei et~al.\ 2013).  This study found that once the various parameters are
estimated for each model individually, the $R_{\rm h}=ct$ cosmology
provides a better fit to the data.  And although about $20\%$ of the GRB
events lie at~least $2\sigma$ away from the best-fit curves (in both
models), suggesting either that some contamination by non-standard GRB
luminosities is unavoidable, or that the errors and intrinsic scatter
associated with the data are being underestimated, various model selection
techniques applied to the GRB data show that the likelihood of $R_{\rm
  h}=ct$ being the correct model rather than $\Lambda$CDM is $\sim 90\%$
versus $\sim 10\%$.

Moreover, although the redshift--distance relationship is essentially the
same in $\Lambda$CDM and the $R_{\rm h}=ct$ Universe (even out to $z\ga
6-7$), the redshift--age relationship is not.  This motivated a recent
examination of whether or not the observed growth rate of high-$z$ quasars
could be used to test the various models.  Quasars at $z\ga 6$ are now
known to be accreting at, or near, their Eddington limit (see, e.g.,
Willott et~al.\ 2010a,b; De~Rosa et~al.\ 2011), which presents a problem
for $\Lambda$CDM because this makes it difficult to understand how $\sim
10^9\;M_\odot$ supermassive black holes could have appeared only 700--900
Myr after the big~bang.  Instead, in $R_{\rm h}=ct$, their emergence at
redshift $\sim 6$ corresponds to a cosmic age of $\ga 1.6$ Gyr, which was
enough time for them to begin growing from $\sim 5-20\; M_\odot$ seeds
(presumably the remnants of Pop~II and~III supernovae) at $z\la 15$ (i.e.,
\emph{after} the onset of re-ionization) and still reach a billion solar
masses by $z\sim 6$ via standard, Eddington-limited accretion (Melia
2013a).

In light of this apparent conflict between the implications of the Type~Ia
supernova work and the results of other studies, we have begun to look more
closely at the possibility of directly comparing how $\Lambda$CDM and
$R_{\rm h}=ct$ account for the supernova measurements themselves.  This is
not an easy task, principally because of the enormous amount of work that
goes into first establishing the SN magnitudes, and then carefully fitting
the data using the comprehensive set of parameters available to
$\Lambda$CDM (see, e.g., Suzuki et~al.\ 2012).  In a previous paper (Melia
2012a), the redshift--distance relationship in $\Lambda$CDM was compared
with that predicted by $R_{\rm h}=ct$, and each with the Union2.1 sample.
It was shown that the two theories produce virtually indistinguishable
profiles, though the fit with $R_{\rm h}=ct$ had not yet been fully
optimized.  That~is, despite the fact that this previous analysis simply
used $R_{\rm h}=ct$ to fit the data optimized with $\Lambda$CDM, the
results were quite promising, suggesting that a full optimization
procedure---followed separately for $R_{\rm h}=ct$ and $\Lambda$CDM---ought
to be carried~out.  The principal goal of this paper is to complete this
study.

A direct comparison of $\Lambda$CDM with $R_{\rm h}=ct$ using Type~Ia SNe
is also motivated by recent theoretical work suggesting that the
Friedmann--Robertson--Walker (FRW) metric is more specialized than was
previously thought (Melia 2013b).  A close examination of the physics
behind the symmetries incorporated into this well-known and often employed
solution to Einstein's equations showed that the FRW metric applies only to
a fluid with zero active mass, i.e., a fluid in which ${\rho+3p=0}$, where
$p$~is the {\it total} pressure and $\rho$ the {\it total} energy density.
This is consistent with the equation-of-state used in $R_{\rm h}=ct$, but
not in $\Lambda$CDM, except that when one averages the pressure, $\langle
p\rangle$, over the age of the Universe, one does get $\approx
-\langle\rho\rangle/3$ for the estimated $\Lambda$CDM parameters.
$\Lambda$CDM therefore appears to be an empirical approximation to $R_{\rm
  h}=ct$, asymptotically approaching the requirements of the zero active
mass condition consistent with the FRW metric.

In \S~2 of this paper, we briefly summarize the contents of the SNLS
sample, and explain the model parameters (including those associated with
the data) that are to be estimated.  We present the fits to the supernova
data in~\S~3, and a direct comparison between $\Lambda$CDM and the $R_{\rm
  h}=ct$ Universe is made in~\S~4.  We~end with a discussion and
conclusions in~\S~5.

\section{The SNLS Supernova Sample\label{sec:intro}}
The Supernova Legacy Survey (SNLS) sample contains 252 high-redshift
(${0.15<z<1.1}$) Type~Ia SNe discovered during the first three years of
operation (Guy et~al.\ 2010).  One of the most important features of this
catalog is that it constitutes a single, homogeneous sample.  It also
covers the very important redshift range where the standard model suggests
the Universe underwent a transition from decelerated to accelerated
expansion.  Guy et~al.\ (2010) used two light-curve fitters (SALT2 and
SiFTO) to determine the peak magnitudes, light-curve shapes, and colors of
the Type~Ia SNe.

For SiFTO, a distance modulus is defined for each SN as the linear
combination
\begin{equation}
\mu_{B}=m_{B}+\alpha \cdot (s-1)-\beta \cdot \mathcal{C}-M_B,
\end{equation}
where $m_{B}$~is the peak rest-frame \emph{B}-band magnitude, $s$~is the
stretch (a measure of light-curve shape), $\mathcal{C}$~is the color (peak
rest-frame ${B-V}$), and $M_B$~is the absolute magnitude of a Type~Ia
supernova.\footnote{The SNLS supernova compilation of 252 SNe is currently
  available in the University of Toronto's Research Repository, at
  https://tspace.library.utoronto.ca/handle/1807/24512.  It includes the
  following information for each~SN: $m_{B}$~(with corresponding standard
  error~$\sigma_{m_{B}}$), $s$~(with error~$\sigma_{s}$),
  $\mathcal{C}$~(with error~$\sigma_{\mathcal{C}}$), and the covariances
  between $m_{B},s,\mathcal{C}$.}  When corrected for shape and color,
Type~Ia SN luminosities have a dispersion of only~$\sim15\%$.  The
coefficients $\alpha,\beta$ are thus the parameters of a luminosity model,
though it is not a full statistical model, since it lacks an explicit error
dispersion parameter.  In the present context, $\alpha,\beta$ and~$M_B$ are
`nuisance' parameters, as they cannot be estimated independently of an
assumed cosmology.  They must be optimized simultaneously with the
cosmological parameters, as will be explained in~\S\,3.

The theoretical distance modulus $\mu_{\rm th}$ is calculated for each~SN
from its measured redshift $z$ by
\begin{equation}
\mu_{\rm th}(z)\equiv5\log(D_{L}(z)/ 10 \rm pc),
\label{distance}
\end{equation}
where $D_{L}(z)$ is the model-dependent luminosity distance.
A~determination of~$D_{L}$ requires the assumption of a particular
expansion scenario.  Both $\Lambda$CDM and $R_{\rm h}=ct$ are FRW
cosmologies, but the former assumes specific constituents in the density,
written as $\rho=\allowbreak\rho_r+\nobreak\rho_m+\nobreak\rho_\Lambda$,
where $\rho_r,\rho_m,\rho_\Lambda$ are, respectively, the energy densities
of radiation, matter (luminous and dark), and the cosmological
constant~$\Lambda$.  These are often expressed in~terms of today's critical
density, $\rho_c\equiv 3c^2 H_0^2/8\pi G$, where $H_0$~is the Hubble
constant, by $\Omega_m\equiv\rho_m/\rho_c$, $\Omega_r\equiv\rho_r/\rho_c$,
and $\Omega_\Lambda\equiv \rho_\Lambda/\rho_c$.  In~a flat universe with
zero spatial curvature, the total scaled energy density is
$\Omega\equiv\allowbreak\Omega_m+\nobreak\Omega_r+\nobreak\Omega_\Lambda\nobreak=\nobreak1$.
Since $\Omega_r\ll1$, $\Omega_m+\nobreak\Omega_\Lambda\nobreak=\nobreak1$.
In $R_{\rm h}=ct$, on the other hand, whatever constituents are present
in~$\rho$, the principal constraint is the total equation-of-state
$p=-\rho/3$, which as we mentioned in~\S\,1, is in~fact required by the use
of the FRW metric.

When dark energy is included with an unknown equation-of-state,
$p_\Lambda=w_\Lambda\rho_\Lambda$, the general form of the luminosity
distance in $\Lambda$CDM is given by
\begin{multline}
D_{L}^{\Lambda {\rm CDM}}(z) = {c\over
  H_{0}}\,{(1+z)\over\sqrt{\mid\Omega_{k}\mid}}\; {\rm
  sinn}\Biggl\{\mid\Omega_{k}\mid^{1/2} \times \\
\int_{0}^{z}{dz\over\sqrt{\Omega_{\rm m}(1+z)^{3}+\Omega_{k}(1+z)^{2}+\Omega_{\Lambda}(1+z)^{3(1+w_\Lambda)}}}\Biggr\}\;,
\end{multline}
where $c$ is the speed of light.  In this, $\Omega_{k}=1-\Omega_{\rm
  m}-\Omega_{\Lambda}$ represents the spatial curvature of the
Universe---appearing as a term proportional to the spatial curvature
constant $k$ in the Friedmann equation.  In addition, sinn is $\sinh$ when
$\Omega_{k}>0$ and $\sin$ when $\Omega_{k}<0$.  For a flat Universe
($\Omega_{k}=0$), the right side becomes $(1+z)c/H_{0}$ times the
indefinite integral.

In the $R_{\rm h}=ct$ Universe, the luminosity distance is given by the
much simpler expression
\begin{equation}
D_{L}^{R_{\rm h}=ct}(z)=\frac{c}{H_{0}}(1+z)\ln(1+z)\;.
\end{equation}
The factor $c/H_0$ is in~fact the gravitational horizon $R_{\rm h}(t_0)$
(which itself is coincident with the Hubble radius) at the present time, so
we may also write the luminosity distance as
\begin{equation}
D_{L}^{R_{\rm h}=ct}(z)=R_{\rm h}(t_0)(1+z)\ln(1+z)\;.
\end{equation}
A more extensive description of the observational differences between
$\Lambda$CDM and $R_{\rm h}=ct$ is provided in Melia (2007,2012a), Melia \&
Shevchuk (2012), Melia \& Maier (2013), and Wei et~al.\ (2013).  For a
pedagogical treatment, see also Melia (2012b).

\section{Theoretical Fits\label{sec:fits}}
In this paper, we shall use two methods for calculating point and interval
estimates of cosmological model parameters: one a method commonly used in
past analyses (e.g., Kowalski et~al.\ 2008; Guy et~al.\ 2010; Suzuki
et~al.\ 2012), and one more recently proposed, which is based on maximizing
the likelihood function (e.g., Kim 2011; Melia \& Maier 2013; Wei
et~al.\ 2013).  As we shall see, the latter allows us to estimate all
parameters, including the unknown intrinsic dispersion.  But we shall find
that under some circumstances, the manner in which this latter method works
offers some justification for the former, which instead, minimizes~$\chi^2$
subject to the condition that the reduced~$\chi^2$ equal unity.  In either
method, however, nuisance parameters characterizing the SN~luminosities,
and free parameters of the cosmological model, must be fitted
simultaneously.

\emph{Initial Remark on Free Parameters.} Before we begin the actual
fitting of the various models, we take a moment to summarize the free
parameters available in each.  As alluded to earlier, it is now understood
that the symmetries incorporated into the FRW metric require a zero active
mass condition (i.e., $\rho+3p=0$) in the cosmic fluid (Melia 2013b).  One
can understand this even without a formal proof, by using the following
reasoning.

In spherically symmetric spacetimes, which include the special FRW case, a
proper mass emerges from the introduction of the metric into Einstein's
field equations.  We call this the Misner--Sharp mass (Misner \& Sharp
1964), which is defined as $M(R_{\rm h})=(4\pi/3)R_{\rm h}^3 \rho/c^2$, in
terms of the gravitational radius $R_{\rm h}=2GM(R_{\rm h})/c^2$.  This
proper mass is a consequence of Birkhoff's theorem (Birkhoff 1923), which
states that none of the mass-energy beyond~$R_{\rm h}$ contributes to the
spacetime curvature within a shell at this radius.  The gravitational
radius must therefore be defined as written here, although the cosmos
itself may be infinite.  But since $M$ is a proper mass defined in~terms of
the proper density~$\rho$ and proper volume, $R_{\rm h}$~must therefore be
a proper radius as~well.  Thus, by Weyl's postulate (Weyl 1923), this
gravitational radius must have the form $R_{\rm h}=ar_{\rm h}$, where
$a(t)$ is the universal expansion factor and $r_{\rm h}$~is a constant
comoving distance.  One can therefore easily show from the first Friedmann
equation (see, e.g., Melia \& Shevchuk 2012) that $\dot{a}^2+kc^2=r_{\rm
  h}^{-2}$, where $k$~is the spatial curvature constant.  In other words,
the expansion rate~$\dot{a}$ must be constant, which then leads, via the
second Friedmann (or `acceleration') equation, to the condition
$p=-\rho/3$.  The equation-of-state $w=-1/3$ is therefore required in this
cosmology, and cannot be adjusted to optimize fits to the data.

Further, since $H\equiv (\dot{a}/a)=1/t$, one can easily see that the
Hubble radius $R_{\rm Hubble}\equiv c/H$ must be given as $R_{\rm
  Hubble}=ct$ in this cosmology.  But in general relativity, we recognize
such a surface moving at proper speed~$c$ relative to an observer in the
co-moving frame as an event horizon.  So if the Misner--Sharp mass gives
rise to the gravitational radius~$R_{\rm h}$, which is identified with
$R_{\rm Hubble}$, then from the Friedmann equation we must also have $k=0$.
The $R_{\rm h}=ct$ Universe is therefore flat and has a unique, fixed
equation-of-state $w=-1/3$.  It~has only one free parameter: the Hubble
constant~$H_0$.

In $\Lambda$CDM there is some flexibility in choosing our free parameters
for the analysis of Type~Ia SNe, depending on how we characterize dark
energy.  At the very minimum, we may adjust $\Omega_{\rm m}$, $k$ (or,
equivalently,~$\Omega_{\rm de}$), $w_{\rm de}$ and~$H_0$.  (Note that
additional free parameters, such as the baryon density, may be relevant to
fits involving the CMB, but are not necessary for constructing the SN
Hubble diagram.)  And though inflation is not yet a fully self-consistent
model, we may include it as a foundation for $\Lambda$CDM, in which case we
could reasonably argue that $k=0$.  However, even inflation does not
specify the equation-of-state for dark energy.  And using priors from other
kinds of observations is not really appropriate for this work, because
their values change from observation to observation---and from instrument
to instrument.  For example, note that the values of $w_{\rm de}$
and~$\Omega_{\rm m}$ from WMAP (Bennett et~al.\ 2003) are quite different
from those inferred by {\it Planck} (Ade et~al.\ 2014).  The most basic
$\Lambda$CDM model we may use for supernova work thus has three free
parameters.  Below, we shall choose the parameters $\Omega_{\rm m}$,
$w_{\rm de}$ and~$H_0$ to represent the standard model, since these are the
three that produce the most favorable fits when using $\Lambda$CDM, based
on what is currently known.

\emph{Method~I.}  The commonly employed fitting method, as discussed by Guy
et~al.\ (2010), determines the best-fit cosmology by (iteratively)
minimizing the $\chi^2$ function
\begin{equation}
\begin{split}
\chi^{2}=\sum_i\frac{\left(\mu_{B,i} - \mu_{\rm th}(z_i)\right)^2}{\sigma^{2}_{{\rm lc},i}+\sigma^{2}_{\rm int}}
\end{split}
\end{equation}
over a joint parameter space, thereby optimizing both the nuisance
parameters $\alpha,\beta,M_B$ and the cosmological parameters.  Here
$i$~ranges over the serial numbers of the SNe in the SNLS catalog, and
$\mu_{B,i}$ depends on~$\alpha,\beta,M_B$ according to Equation~1; and
of~course, $\mu_{\rm th}(z_i)$ depends on the cosmological parameters.  The
SN-specific dispersion $\sigma^2_{{\rm lc},i}$ is defined by
\begin{equation}
\sigma^{2}_{{\rm lc},i}=\sigma^{2}_{m_{B},i}+\alpha^{2}\sigma^{2}_{s,i}+
\beta^{2}\sigma^{2}_{\mathcal{C},i}+C_{m_{B}\,s\,\mathcal{C},i}\;,
\end{equation}
where $\sigma_{{m_{B}},i}$, $\sigma_{{s},i}$, and
$\sigma_{{\mathcal{C}},i}$ are the standard errors of the peak magnitude
and light-curve parameters of the~SN\null.  The term
$C_{m_{B}\,s\,\mathcal{C},i}$ comes from the covariances among
$m_{B},s,\mathcal{C}$, and likewise depends quadratically
on~$\alpha,\beta$.  The quantity $\sigma_{\rm int}$ is an intrinsic (i.e.,
SN-independent) dispersion, the value of which is set by requiring that the
reduced $\chi^{2}$ equal unity.  This additional dispersion takes into
account, e.g., the residual ($\sim15\%$) variability in the Type~Ia SN
luminosity, not captured by the correction coefficients~$\alpha,\beta$.

The following is how \emph{Method~I} is applied, yielding fitted values for
the cosmological parameters (with the exception of the Hubble
constant~$H_0$---more on this below), and model-specific fitted values for
the nuisance parameters $\alpha,\beta,M_B$.  First, $\sigma_{\rm int}$~is
initialized to zero.  Second, the $\chi^2$~function is minimized over the
cosmological and nuisance parameters, and their values are updated.  Third,
with these updated values fixed, the value of~$\sigma_{\rm int}$ for which
the reduced~$\chi^2$, i.e., $\chi^2/{\rm dof}\equiv \chi^2/(n-k)$, equals
unity is determined, and $\sigma_{\rm int}$~is updated.  (Here, $n$
[$=234$, see below] is the number of SNe and $k$~is the total number of
parameters, including~$\sigma_{\rm int}$.)  Steps 2 and~3 are repeated
until the parameters converge within tolerance to stable values.

To find the best-fit coefficients $\alpha$, $\beta$, $M_{B}$ and the
cosmological parameters that define the fitted model, we use Markov-chain
Monte Carlo (MCMC) techniques in our calculations. Our MCMC approach
generates a chain of sample points distributed in the parameter space
according to the posterior probability, using the Metropolis-Hastings algorithm
with uniform prior probability distributions, such that $0.5<\alpha<2.0$,
$2.0<\beta<4.0$, $-19.5<M_{B}<-18.5$, $0.0<\sigma_{\rm int}<1.0$,\footnote{In
\emph{Method~II}, $\sigma_{\rm int}$~is treated on the same level as
the other parameters---more on this below.} $0.0<\Omega_{\rm m}<1.0$,
and/or $0.0<\Omega_{\Lambda}<1.8$ (in the non-flat case). In the
parameter space formed by the constrained nuisance parameters and
cosmological parameters, a random set of initial values of the model
parameters is chosen to calculate the $\chi^{2}$  (\emph{Method~I}),
or the likelihood function (\emph{Method~II}). Whether the set of
parameters can be accepted as an effective Markov chain or not is
determined by the Metropolis-Hastings algorithm. The accepted set
not only forms a Markov chain, but also provides a starting point for
the next process. We then repeat this process until the established
convergence accuracy can be satisfied. For each Markov chain, we
generate $10^{5}$ samples consistent with the likelihood function.
Then we derive the probability distributions of the coefficients
$\alpha$, $\beta$, $M_{B}$, and the cosmological parameters
from a statistical analysis of the samples.

To be consistent with previous work (Guy et~al.\ 2010), we discard from the
SNLS catalog all SNe with a peak rest-frame $(B-V)>0.2$.  Such red SNe are
found only at $z<0.6$ in the SNLS because they are fainter than the
average, and hence are undetected (or unidentified spectroscopically) at
higher redshifts.  Discarding them minimizes any potential biasing of the
distance modulus by an inadequate color correction.  Indeed, the correction
coefficient~($\beta$) we estimate from the bulk of the SNe may not apply to
those red SNe that are more likely to be extinguished by dust in their host
galaxy than bluer ones.  This cut, applied to both SALT2 and SiFTO samples,
discards 11~SNe.  There are also 3~SNe, the peak magnitudes of which could
not be obtained with SiFTO due to a lack of observations in the $g_{M}$ and
$z_{M}$~bands.  Finally, again following Guy et~al.\ (2010), we remove SNe
that are $3\sigma$~outliers for either of the best-fit $\Lambda$CDM and
$R_{\rm h}=ct$ models.  This removes 4~more SNe from both the $\Lambda$CDM
and $R_{\rm h}=ct$ samples.  Three (o4D3dd, 04D1ak, and 04D4gz) are in
common; 03D4cx is removed for $\Lambda$CDM, while 05D2ei is removed for
$R_{\rm h}=ct$.  In total, that leaves $n=234$ SNLS supernovae for the
analysis.

Although the number of free parameters in the dark-energy model can be as
large as eight, depending on how one handles the dark energy, in this paper
we take a minimalist approach and use only three of the most essential
ones; these are taken from among a set that includes the Hubble
constant~$H_{0}$, the matter energy density~$\Omega_{\rm m}$, the dark
energy density~$\Omega_{\Lambda}$, and the dark energy equation-of-state
parameter~$w_\Lambda$.  In fitting, the chosen value of~$H_{0}$ is not
independent of~$M_{B}$.  That~is, one can vary either of $H_0,M_B$, but not
both.  Therefore, if we take $M_{B}$ to be a member of the set of nuisance
parameters characterizing the SN luminosities, the dark-energy model will
have, at most, only two parameters: $\Omega_{\rm m}$
and~$\Omega_{\Lambda}$.  For consistency with Guy et~al.\ (2010), we use
$H_{0}=70$ km $\rm s^{-1}$ $\rm Mpc^{-1}$ (recalling that this value does
not actually affect any of the fits).

\begin{figure}[hp]
\centerline{\includegraphics[angle=0,scale=1.0]{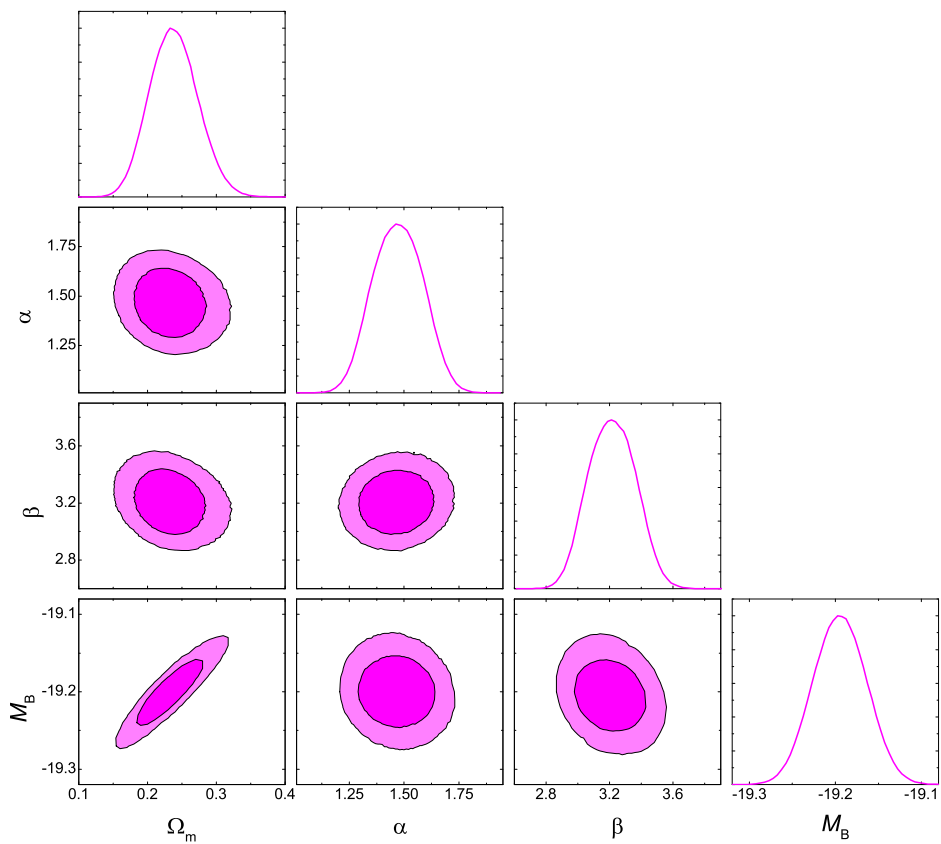}}
\vskip -0.4in
\caption{The (normalized) likelihood distributions and 2-D joint
  distributions with $1\sigma,2\sigma$ contours, for the coefficients
  $\alpha,\beta,M_B$ in Equation~1 and the matter density
  parameter~$\Omega_{\rm m}$, obtained from a constrained minimization
  of~$\chi^2$ (\emph{Method~I}), with the assumption of flatness
  ($\Omega_\Lambda =\allowbreak1-\nobreak\Omega_{\rm m}$).
  The $1\sigma$ confidence intervals are well approximated by Gaussians.}\label{test1}
\end{figure}

\begin{figure}[hp]
\centerline{\includegraphics[angle=0,scale=1.2]{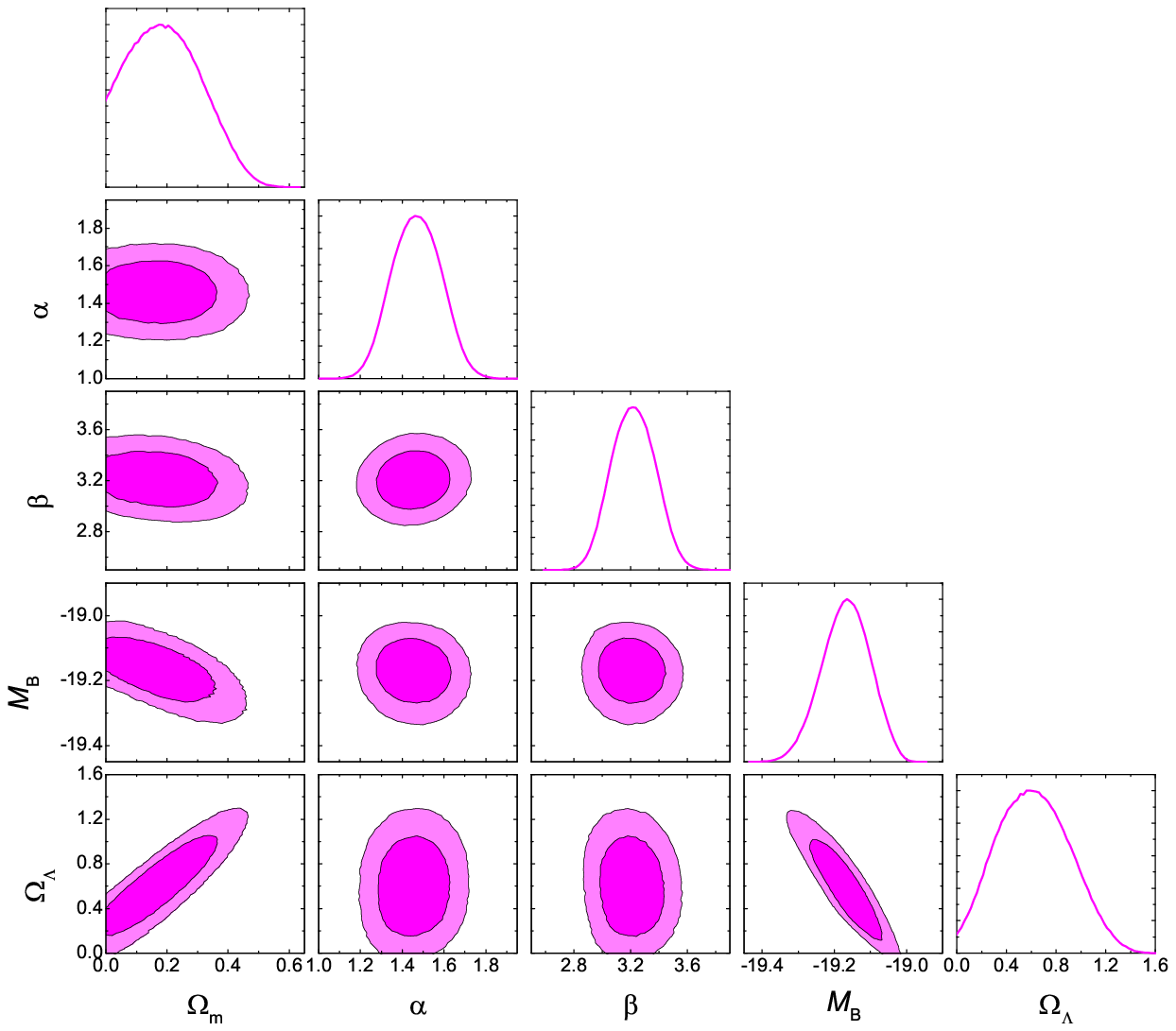}}
\vskip -0.4in
\caption{The (normalized) likelihood distributions and 2-D joint
  distributions with $1\sigma,2\sigma$ contours, for the coefficients
  $\alpha,\beta,M_B$ in Equation~1 and the cosmological parameters
  $\Omega_{\rm m},\Omega_{\Lambda}$, obtained from a constrained
  minimization of~$\chi^{2}$ (\emph{Method~I}), without the assumption of
  flatness.}\label{test2}
\end{figure}

To ensure that our implementation of \emph{Method~I} is reliable and
consistent with the results of Guy et~al.\ (2010), we first examine a
simple test case: the flat Universe, in which
$\Omega_{\Lambda}=\allowbreak1-\nobreak\Omega_{\rm m}$, with
equation-of-state parameter $w_\Lambda=-1$.  $\Omega_{\rm m}$~is then the
only free cosmological parameter.  Guy et~al.\ (2010) found that the
assumption of flatness resulted in a value for $\Omega_{\rm m}$ that we now
know is inconsistent with that reported by Planck2013 (Ade et~al.\ 2014).
Nonetheless, it is useful to begin with this test case because it allows us
to compare our results directly with theirs.  Applying the above
$\chi^2$-minimization procedure, we find that the fitted parameters in this
case are $\alpha=1.470\pm0.121$, $\beta=3.207\pm0.157$,
$M_B=-19.197\pm0.031$, and $\Omega_{\rm m}=0.235\pm0.035$, with
$\sigma_{\rm int}=0.090$.  In Figure~1 we show the (normalized) likelihood
distribution for each parameter ($\alpha,\beta,M_B;\allowbreak\Omega_{\rm
  m}$), according to the factor $\exp(-\chi^2/2)$, and $1\sigma$,$2\sigma$
contours for the joint distribution of each pair of parameters.  For each
parameter, the likelihood distribution is well approximated by a Gaussian,
and the stated confidence interval is a~$68\%$ (i.e.,~$\pm1\sigma$)
interval for this Gaussian.  Using the same method and conditions, Guy
et~al.\ (2010) obtained $\alpha=1.487\pm0.097$, $\beta=3.212\pm0.128$,
$M_B=-19.210\pm0.030$, and $\Omega_{\rm m}=0.215\pm0.033$, with
$\sigma_{\rm int}=0.087$.  These results are quite consistent, again
attesting to the reliability of our calculation.

To allow for a more realistic fit to the SNLS supernova data, we removed
the flatness assumption, and allowed both $\Omega_{\rm m}$
and~$\Omega_{\Lambda}$ to be free parameters.  By minimizing~$\chi^2$, we
now obtain $\alpha=1.469\pm0.123$, $\beta=3.209\pm0.159$,
$M_B=-19.187\pm0.068$, and $\Omega_{\rm m}=0.217\pm0.150$,
$\Omega_{\Lambda}=0.718\pm0.329$, with $\sigma_{\rm int}=0.090$.  The
(normalized) likelihood distribution for each parameter
($\alpha,\beta,M_B;\allowbreak\Omega_{\rm m},\Omega_{\Lambda}$), and a
contour plot for each two-parameter combination, are shown in Figure~2.
Dropping the flatness condition, i.e., fitting both $\Omega_m$
and~$\Omega_\Lambda$, makes the best-fit $\Lambda$CDM model marginally
consistent with the Planck2013 results, because the $1\sigma$ standard
errors are now considerably larger than in the previous case.

It should be noted that the $\chi^{2}$ approach of \emph{Method~I} does not
actually maximize a likelihood.  Rather, it calculates a value
for~$\sigma_{\rm int}$ (with no accompanying uncertainty) by requiring that
$\chi^{2}_{\rm dof}$ equal unity.  Driving $\chi^{2}_{\rm dof}$ to unity,
irrespective of how well the cosmological model fits the data, makes it
impossible to perform a fair comparison of competing models by using
\emph{Method~I}, especially when they have different numbers of parameters.
A~statistically valid analysis must estimate error dispersion parameter(s),
along with all other parameters, by maximizing a joint likelihood function.

\emph{Method~II.}  As a superior alternative to \emph{Model I} for
estimating cosmological parameters, and (simultaneously) the model-specific
optimized nuisance parameters $\alpha,\beta,M_B$, we employ a method
described in D'Agostini (2005) and Kim (2011), which is based on the
maximization of a joint likelihood function.  The joint likelihood function
for all these parameters and the intrinsic dispersion $\sigma_{\rm int}$,
based on a flat Bayesian prior, is
\begin{equation}
L = \prod_{i}
\frac{1}{\sqrt{2\pi(\sigma^{2}_{{\rm lc},i}+\sigma^{2}_{\rm int})}}\;\times
\exp\left[-\,\frac{\left(\mu_{B,i}-\mu_{\rm
      th}(z_i)\right)^{2}}{2(\sigma^{2}_{{\rm lc},i}+\sigma^{2}_{\rm
      int})}\right] \propto \exp\left(-\chi^2/2\right).
\end{equation}
As in \emph{Method~I}, each distance modulus~$\mu_{B,i}$ depends on
$\alpha,\beta,M_B$, and the theoretical distance modulus $\mu_{{\rm
    th}}(z_i)$ depends on the cosmological parameters.  Maximum likelihood
estimation (MLE) is, of~course, a standard statistical procedure, and
appeared in \emph{Method~I} in a limited way, as a minimization
of~$\chi^2$.  The new feature in \emph{Method~II} is that $\sigma_{\rm
  int}$~is treated on the same level as the other parameters: the
likelihood~$L$ is maximized over all parameters, now including~$\sigma_{\rm
  int}$.  This `full~MLE' provides a statistically founded method for
estimating~$\sigma_{\rm int}$ (Kim 2011).  It also treats on the same level
the cosmological parameter uncertainties and the potential uncertainty
in~$\sigma_{\rm int}$, which can affect each other.

The optimized $\Lambda$CDM parameters obtained by \emph{Method~II} are:
$\alpha=1.275\pm0.120$, $\beta=2.637\pm0.155$, $M_B=-19.165\pm0.081$,
$\Omega_{\rm m}=0.365\pm0.137$, and $\Omega_{\Lambda}=0.846\pm0.353$, with
$\sigma_{\rm int}=0.103\pm0.010$.  For each parameter, the likelihood
distribution is well approximated by a Gaussian, and the stated confidence
interval is a~$68\%$ (i.e.,~$\pm1\sigma$) interval for this Gaussian.  As
anticipated by Kim (2011), these best-fit values are not all consistent
with those of \emph{Method I}, though the estimated value of $\Omega_{\rm
  m}$ and its error are entirely consistent with the Planck2013 results.
In an analysis using mock samples, Kim concluded that differences such as
these arise because the commonly used method based on the constrained
minimization of~$\chi^2$ (i.e., \emph{Method~I}) does not include in its
error propagation the covariance of~$\sigma_{\rm int}$ with the other
parameters.

The \emph{Method~II} values of the coefficients $\alpha,\beta,M_B$ may now
be used to calculate the distance modulus $\mu_{B}$ of Equation~1, for
each~SN\null.  From the distance moduli, we construct the Hubble diagram
shown in the upper panel of Figure~3.  We also show the corresponding SNLS
sample of 234 Type~Ia SNe.  As is now well known, the theoretical fit is
excellent.  The maximum value of the joint likelihood function is given by
$-2\ln L=-238.40$, which we shall need when comparing models using the
Bayes Information Criterion (see below).  For completeness, Figure~3 also
shows the Hubble diagram residuals corresponding to the best-fit
$\Lambda$CDM model. The likelihood distribution obtained by
\emph{Method~II} for each parameter
($\alpha,\beta,M_B;\allowbreak\Omega_{\rm m},\Omega_{\Lambda};\sigma_{\rm
  int}$), and a contour plot of the joint distribution for each
two-parameter combination, are shown in the lower panel.

\begin{figure*}[hp]
\vskip -0.8in
\centerline{\includegraphics[angle=0,scale=0.6]{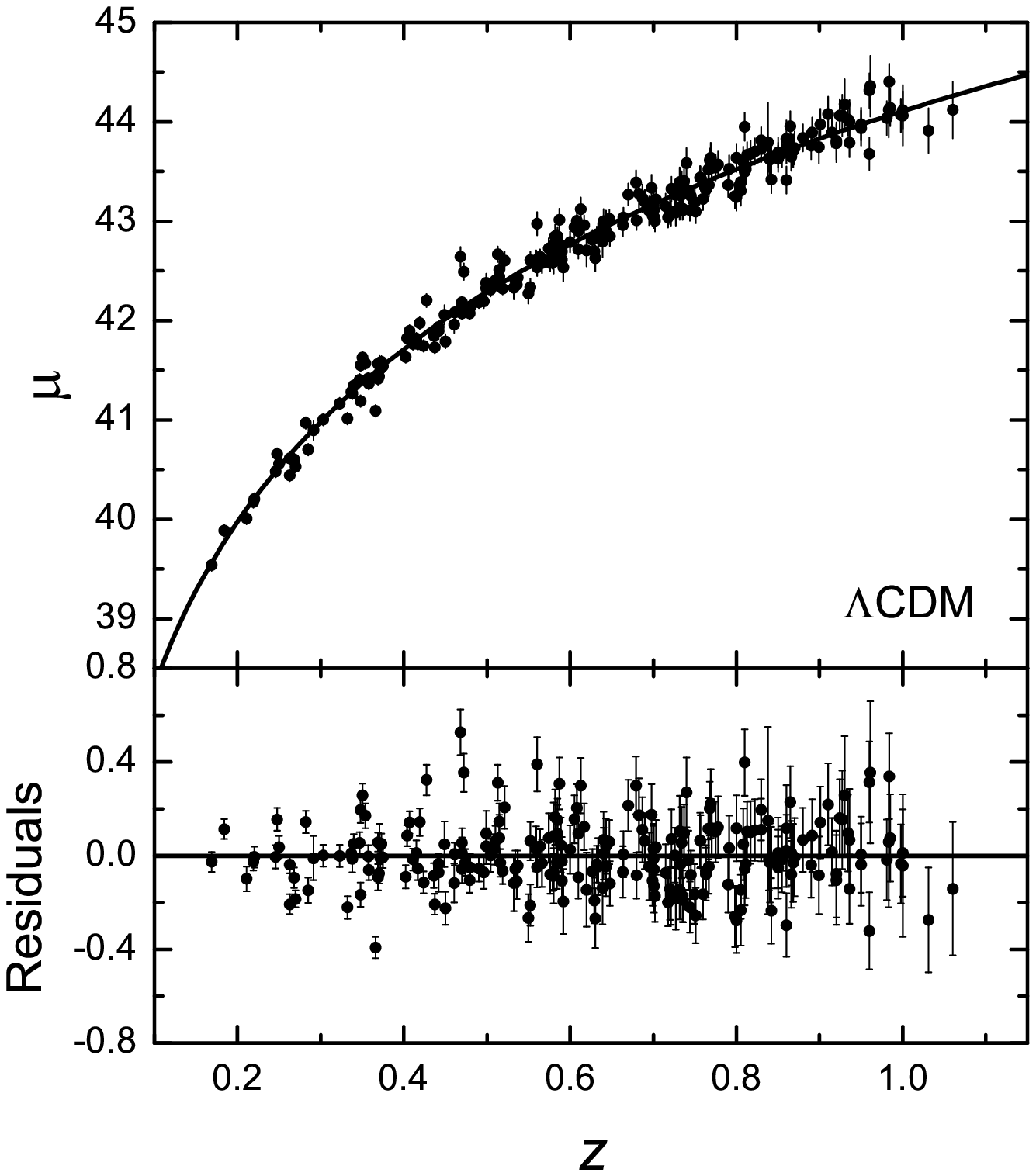}}
\vskip -2.0in
\centerline{\includegraphics[angle=0,scale=1.16]{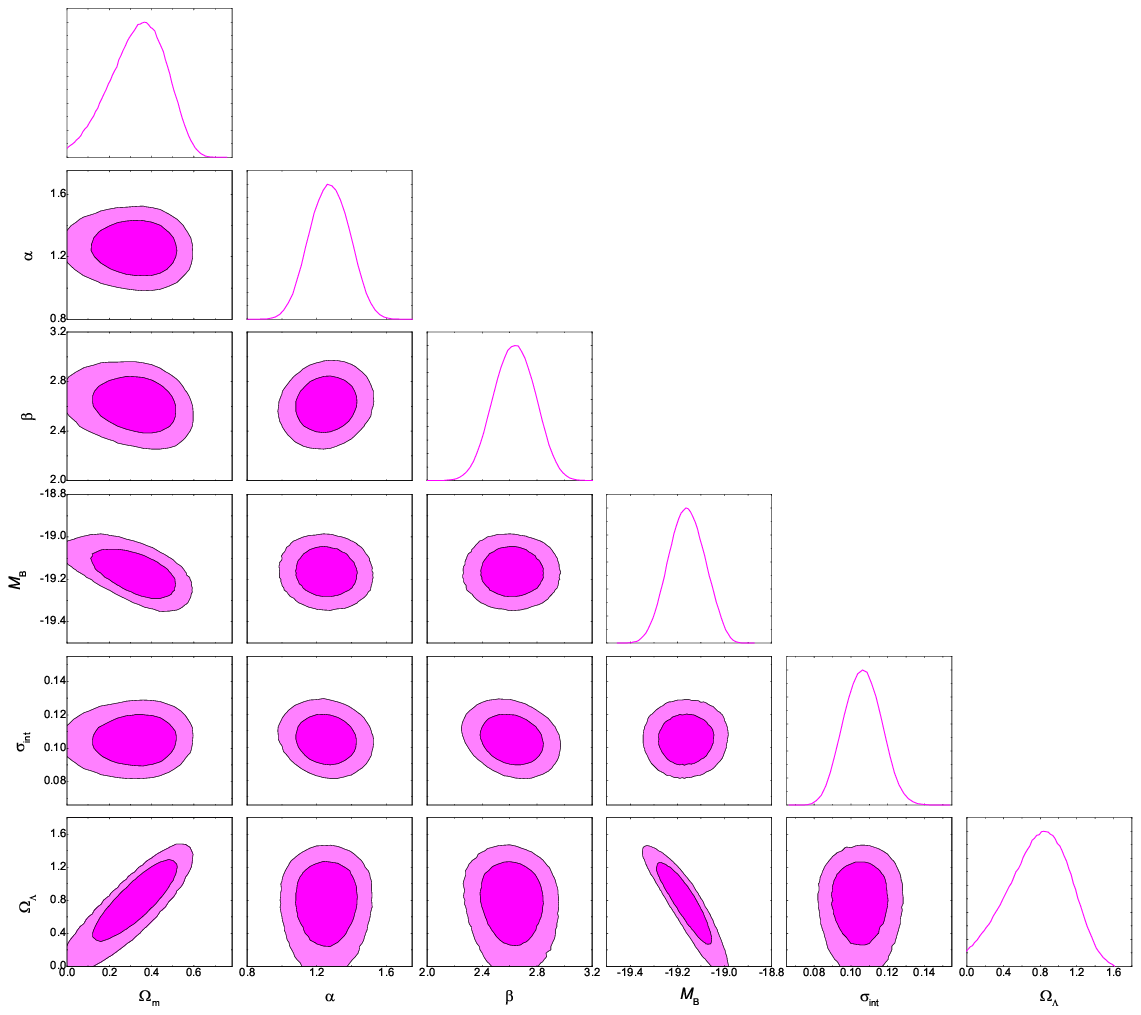}}
\vskip -1.2in
\caption{Upper panel: Hubble diagram and Hubble diagram residuals for the
  SNLS sample of 234 Type~Ia SNe in $\Lambda$CDM\null.  Solid curve:
  best-fit model with $\Omega_{\rm m}=0.365$, $\Omega_{\Lambda}=0.846$
  and $p_\Lambda= -\rho_\Lambda$.  Lower panel:
  (normalized) likelihood distributions and 2-D joint distributions with
  $1\sigma,2\sigma$ contours, for the coefficients $\alpha,\beta,M_B$, the
  intrinsic dispersion $\sigma_{\rm int}$, and $\Omega_{\rm m},
  \Omega_{\Lambda}$.  The fitting method employed is MLE
  (\emph{Method~II}).}\label{LCDM}
\end{figure*}

\section{A Direct Comparison Between $\Lambda$CDM and the $R_{\rm h}=ct$ Universe\label{sec:HD}}
In the $R_{\rm h}=ct$ Universe, there is only one free parameter: the
Hubble constant $H_{0}$.  But since we cannot vary $M_{B}$ and~$H_{0}$
separately, there are no free parameters left to adjust the theoretical
curve, once we optimize $M_{B}$ as one of the nuisance parameters
characterizing the supernova luminosities.

\begin{deluxetable}{lcccccccc}
\tablewidth{465pt}
\tabletypesize{\tiny}
%\tabletypesize{\footnotesize}
\tablecaption{Best Fits for Different Cosmological Models}\tablenum{1}
\tablehead{Model&\colhead{$\alpha$}&\colhead{$\beta$}&\colhead{$M_{B}$}&\colhead{$\sigma_{\rm int}$}
&\colhead{$\Omega_{\rm m}$}&\colhead{$\Omega_{\Lambda}$}& \colhead{$-2\ln L$}&\colhead{BIC} }
\startdata
\emph{Method I} &&&&&&&& \\
$\Lambda$CDM ($k=0$) & $1.470\pm0.121$ & $3.207\pm0.157$ & $-19.197\pm0.031$ &
$0.090\qquad\quad\;$ & $0.235\pm0.035$ & $1-\Omega_{\rm m}\qquad\;\;$ & --- & --- \\
$\Lambda$CDM & $1.469\pm0.123$ & $3.209\pm0.159$ & $-19.187\pm0.068$ &
$0.090\qquad\quad\;$ & $0.217\pm0.150$ & $0.718\pm0.329$ & --- & --- \\
\emph{Method II} &&&&&&&& \\
$\Lambda$CDM   &  $1.275\pm0.120$	& $2.637\pm0.155$ & $-19.165\pm0.081$ & $0.103\pm0.010$  & $0.365\pm0.137$  &  $0.846\pm0.353$    &  -238.40        &   -205.67   \\
$R_{\rm h}=ct$ &  $1.175\pm0.115$	& $2.608\pm0.149$ & $-18.959\pm0.011$ &	$0.106\pm0.010$  &	---      & ---                    &  -231.85        &   -210.03   \\
\enddata
\end{deluxetable}

\begin{figure*}[hp]
\vskip -0.3in
\centerline{\includegraphics[angle=0,scale=0.65]{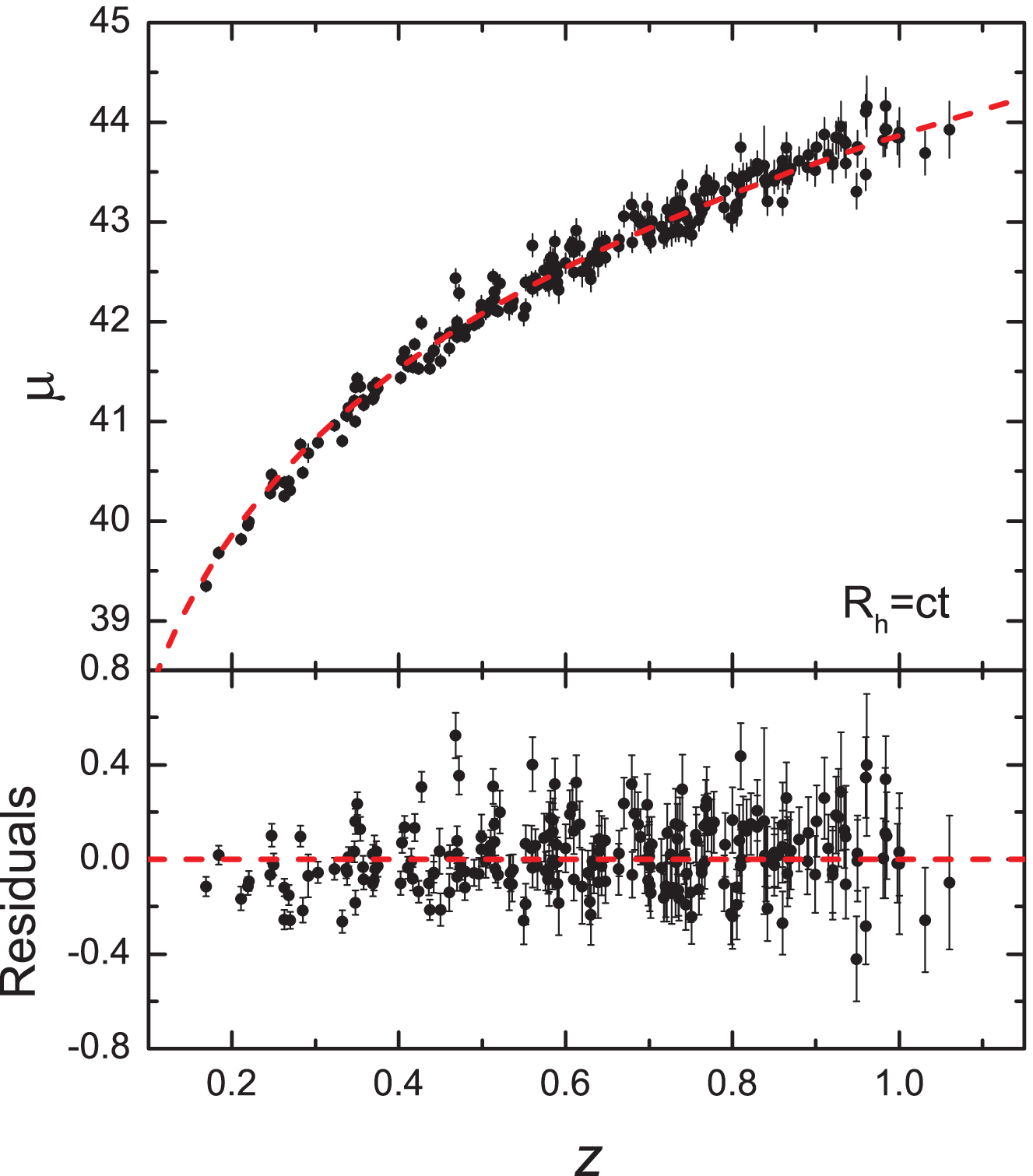}}
\vskip -0.6in
\centerline{\includegraphics[angle=0,scale=0.95]{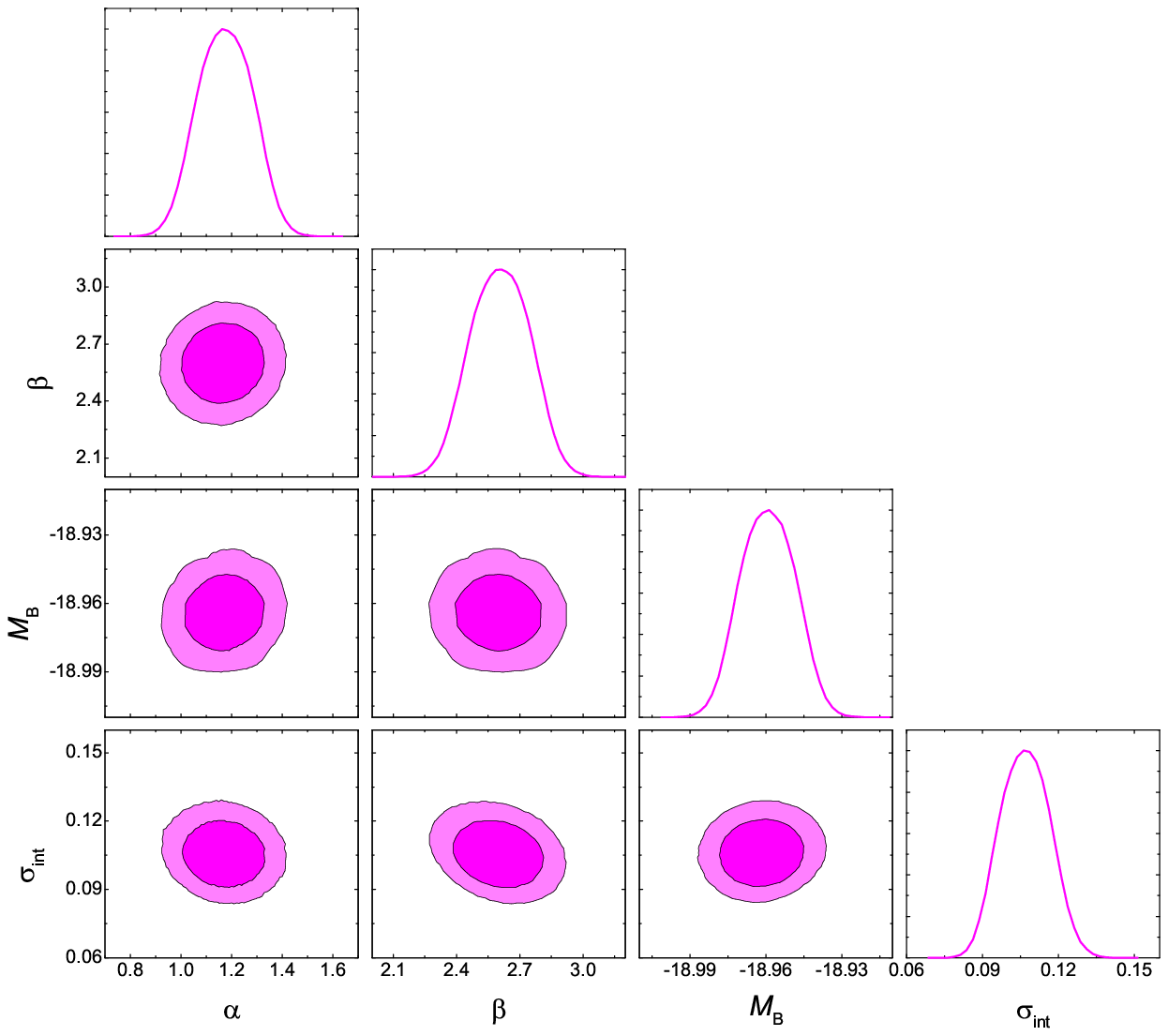}}
\vskip-0.7in
\caption{Same as Figure~3, but now for the $R_{\rm h}=ct$ Universe.
  Upper panel: Hubble diagram and Hubble diagram residuals for the data
  optimized for $R_{\rm h}=ct$, and the corresponding theoretical curve.
  Lower panel: The (normalized) likelihood distributions and 2-D joint
  distributions with $1\sigma,2\sigma$ contours, for all fitted parameters.
  The fitting method employed is MLE (\emph{Method~II}).}\label{Rh}
\end{figure*}

Using the MLE approach (\emph{Method~II}), we find that for the $R_{\rm
  h}=ct$ Universe, the optimized nuisance parameters are
$\alpha=1.175\pm0.115$, $\beta=2.608\pm0.149$, $M_B=-18.959\pm0.011$, with
$\sigma_{\rm int}=0.106\pm0.010$.  All likelihood distributions are well
approximated by Gaussians, and the given confidence intervals are $68\%$
(i.e.,~$\pm1\sigma$) intervals for the Gaussians.  As with $\Lambda$CDM, we
plot the (normalized) likelihood distribution for each parameter
($\alpha,\beta,M_B;\sigma_{\rm int}$), and 2-D plots for two-parameter
combinations (Figure~4).  The best-fit values are quite similar to those
for $\Lambda$CDM, but are not exactly the same, reaffirming the importance
of reducing the data separately for each model being tested.  The $R_{\rm
  h}=ct$ distance modulus is compared to the SNLS sample in the upper panel
of Figure~4.  For completeness, we also show the Hubble diagram residuals
corresponding to the $R_{\rm h}=ct$ Universe at the bottom of this panel.
The maximum value of the joint likelihood function for the optimized
$R_{\rm h}=ct$ fit corresponds to $-2\ln L=-231.85$.  All the fits
performed in this paper are summarized in Table~1, for ease of comparison.

An inspection of the Hubble diagrams in Figures 3 and~4 reveals that the
distance moduli are slightly different when the nuisance parameters are
optimized using different models, but both $\Lambda$CDM and $R_{\rm h}=ct$
fit their respective data sets very well.  One certainly gets this
impression from a side-by-side comparison of the Hubble-diagram residuals
for the SNLS sample in $\Lambda$CDM and $R_{\rm h}=ct$, shown in Figure~5.
However, because these models formulate their observables (such as the
luminosity distance in Equations 3 and~5) differently, and because they do
not have the same number of free parameters, a decision between the models
must be based on a formal model selection technique, and in this regard,
the results of our analysis favor $R_{\rm h}=ct$ over $\Lambda$CDM, as we
shall now demonstrate quantitatively.

A companion paper (Melia \& Maier 2013) discussed at length how one may use
state-of-the art model selection tools to choose the model to be preferred
in accounting for the data.  We shall not reproduce that discussion here,
but we do point~out that to assess competing models in cosmology, a strong
case for using the Bayes Information Criterion (BIC) has been made (see,
e.g., Liddle 2004, 2007; Liddle et~al.\ 2006).  The BIC is applicable when
data points are independent and identically distributed, which is a
reasonable assumption for supernova redshift--luminosity data.  The method
has now been used to compare several popular models against $\Lambda$CDM
(see, e.g., Shi et~al.\ 2012).

\begin{figure}[ph]
\centerline{\hskip0.3in\includegraphics[angle=0,scale=1.25]{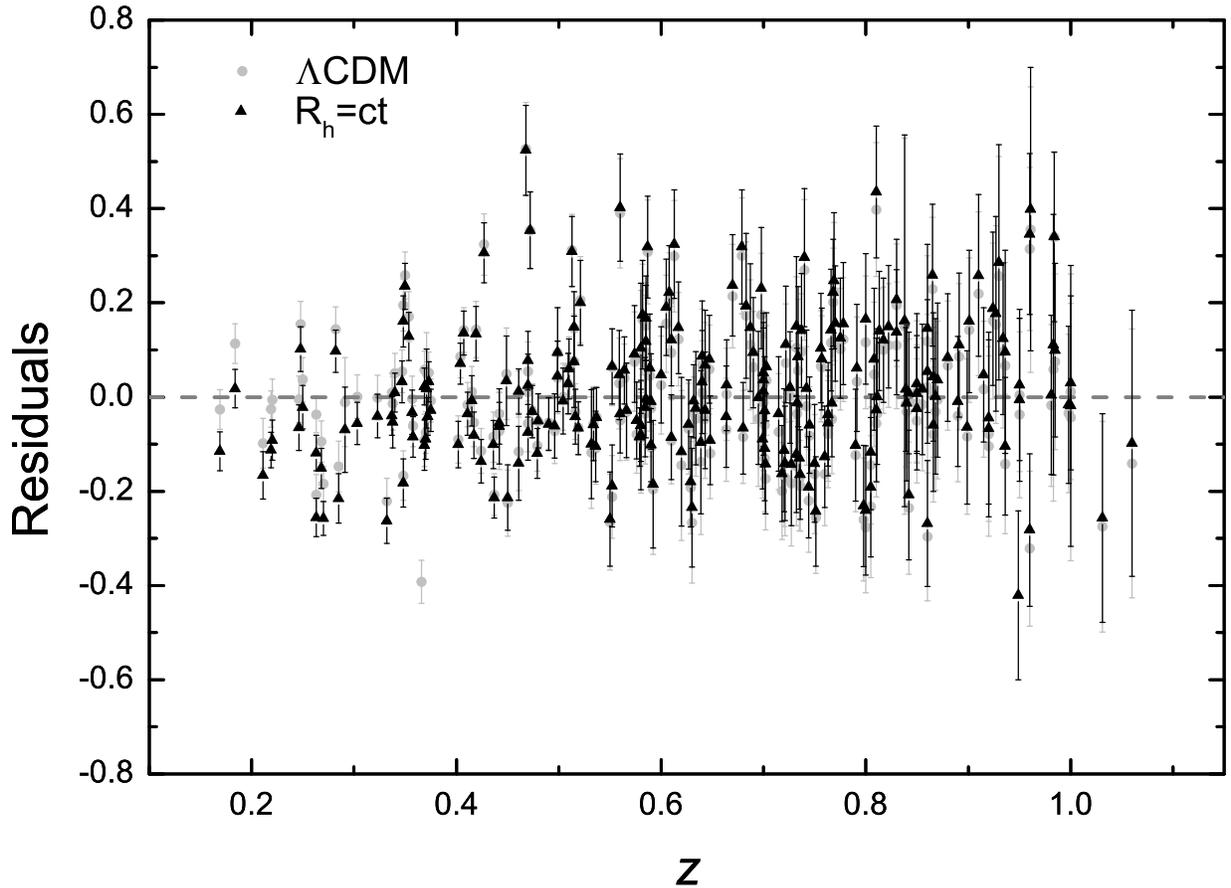}}
%\vskip -1.0in
\caption{A side-by-side comparison of the Hubble-diagram residuals for the
  SNLS sample in the standard ($\Lambda$CDM) model and the $R_{\rm h}=ct$
  Universe.}\label{comparison}
\end{figure}

Despite its name, the BIC is based not on information theory, but rather on
an asymptotic ($n\to\infty$, where $n$~is the number of data points)
approximation to the outcome of a Bayesian inference procedure for deciding
between competing models (Schwarz 1978).  A~comparison between the BIC
values of two or more fitted models provides their relative ranks, and also
a numerical measure of confidence (a likelihood or posterior probability)
that each model is the best.  Unlike some model selection techniques, the
BIC can be applied to `non-nested' models, such as those we have here.  The
BIC for a fitted regression model, linear or nonlinear, captures the
dominant (in the $n\to\infty$ limit) behavior of the associated Bayes
factor, which assesses the strength of the evidence in its favor (Kass \&
Raftery 1995, \S\,4.1).  In this limit, it is only subdominant terms that
are affected by such considerations as the choice of a Bayesian prior (Kuha
2004), or the extent of the model nonlinearity if~any (Haughton 1988).
There is accordingly a simple formula for the BIC, namely
\begin{equation}
{\rm BIC}=-2\ln L+(\ln n)k\;,
\end{equation}
where $k$ is the number of fitted parameters.  The logarithmic penalty term
in the BIC strongly suppresses overfitting if $n$~is large (the situation
we have here, with $n=234$, which is deep in the asymptotic regime).

A quantitative ranking of fitted models 1 and~2 is computed as follows.  If
${\rm BIC}_\alpha$ comes from model~$\alpha$, the unnormalized confidence
in this model is the `Bayes weight' $\exp(-{\rm BIC}_\alpha/2)$.  That is,
in the light of the data, model~$\alpha$ has likelihood
\begin{equation}
P(\alpha)=
\frac{\exp(-{\rm BIC}_\alpha/2)}
{\exp(-{\rm BIC}_1/2)+\exp(-{\rm BIC}_2/2)}
\end{equation}
of being the correct choice.  The strength of the evidence for model~1 and
against model~2 is quantified by $\Delta{\rm BIC}\equiv\allowbreak {\rm
  BIC}_2 -\nobreak {\rm BIC}_1$, and the following qualitative
interpretation of $\Delta{\rm BIC}$ is standard.  If $\Delta{\rm BIC}$ is
less than~$2$, the evidence is ``not worth more than a bare mention'' (Kass
\& Raftery 1995).  If it is in the range $2\dots6$, the evidence is
positive; if it is in the range $6\dots10$, the evidence is strong; and if
it is greater than~$10$, the evidence is very strong.

With $n=234$ data points and $k=4$ parameters, the BIC for the optimized
$R_{\rm h}=ct$ Universe is ${\rm BIC}_{1}=-210.03$.  For the optimized
$\Lambda$CDM, with $k=6$, the corresponding value is ${\rm
  BIC}_{2}=-205.67$.  Our analysis therefore shows that the evidence
supplied by the SNLS sample for the $R_{\rm h}=ct$ Universe over
$\Lambda$CDM is positive; and quantitatively, $R_{\rm h}=ct$ is favored
over $\Lambda$CDM with a likelihood of $\sim 90\%$ versus only~$\sim 10\%$.

\section{Discussion and Conclusions\label{sec:disc}}
Our comparative analysis of $\Lambda$CDM and the $R_{\rm h}=ct$ Universe
using the SNLS sample has shown that---contrary to earlier claims---the
Type~Ia SNe do not point to an expansion history of the Universe in
conflict with that implied by other kinds of source, such as the cosmic
chronometers (Jimenez \& Loeb 2002; Simon et~al.\ 2005; Stern et~al.\ 2010;
Moresco et~al.\ 2012; Melia \& Maier 2013), GRBs (Norris et~al.\ 2000;
Amati et~al.\ 2002; Schaefer 2003; Wei \& Gao 2003; Yonetoku et~al.\ 2004;
Ghirlanda et~al.\  2004; Liang \& Zhang 2005; Liang et~al.\ 2008; Wang
et~al.\ 2011; Wei et~al.\ 2013), and high-$z$ quasars (Kauffmann \&
Haehnelt 2000; Wyithe \& Loeb 2003; Hopkins et~al.\ 2005; Croton
et~al.\ 2006; Fan 2006; Willott et~al.\ 2003; Jiang et~al.\ 2007; Kurk
et~al.\ 2007; Tanaka \& Haiman 2009; Lippai et~al.\ 2009; Hirschmann
et~al.\ 2010; Melia 2013a).  A~difficulty with the use of supernova data has
always been their dependence on the underlying cosmology.  There is no
question that to compare different models properly, one must optimize the
nuisance parameters describing the supernova luminosities separately for
each expansion scenario.  It is not appropriate to use data optimized for
$\Lambda$CDM to test other models.

This has been the primary focus of our analysis in this paper.  We have
confirmed the argument made by Kim (2011), in particular, that one should
optimize parameters by carrying~out a maximum likelihood estimation in any
situation where the parameters include an unknown intrinsic dispersion.
The commonly used method, which estimates the dispersion by requiring the
reduced~$\chi^2$ to equal unity, does not take into account all possible
covariances among the parameters. In this regard, our best-fit models
for $\Lambda$CDM do not agree exactly with those of Guy et al. (2010), who
optimized the model parameters using only \emph{Method I}. Indeed, while
their best-fit model is characterized by parameters noticeably different
from those of Planck2013, we have demonstrated that the use of \emph{Method II}
actually results in an optimized $\Lambda$CDM model consistent with that
based on the analysis of the CMB fluctuations. Simply based on a consideration
of the standard model, therefore, our results support the proposal made by Kim (2011)
that \emph{Method II} should be preferred over \emph{Method I} in the
analysis of Type Ia SN samples that include unknown intrinsic dispersions.

More importantly, we have found that, when the parameter optimization is
handled via the joint likelihood function, both $\Lambda$CDM and $R_{\rm
  h}=ct$ fit their individually optimized data very well.  However, the
$R_{\rm h}=ct$ Universe has only one free parameter---the Hubble
constant~$H_0$---which enhances the significance of the fit over that using
$\Lambda$CDM\null.  Indeed, standard model selection techniques penalize
models with a large number of free parameters.  We have found that the BIC
favors $R_{\rm h}=ct$ over $\Lambda$CDM, by a likelihood of $\sim 90\%$
versus~$\sim 10\%$.  The difference in likelihoods would be even greater
for other variations of $\Lambda$CDM that include a larger number of free
parameters, e.g., to characterize the dark-energy equation-of-state, should
$w_\Lambda$ not be constant.

This result would be quite significant on its own. But when we consider it
in concert with the analysis of all the other data sets that have been
analyzed thus far, it is reasonable to conclude that $R_{\rm h}=ct$ is a
more accurate representation of the Universe than is $\Lambda$CDM\null.
This has far-reaching consequences that will be addressed at greater length
elsewhere.

Obviously, these results call into question the conclusion that the
Universe is currently undergoing a period of acceleration, following an
earlier period of deceleration.  The fact that the fits to the data using
$\Lambda$CDM often come very close to those of $R_{\rm h}=ct$ (see
Figure~5) lends weight to our suspicion that the standard model functions
as an empirical approximation to the latter, since it has more free
parameters and lacks that essential ingredient in $R_{\rm h}=ct$: the
equation-of-state $p=-\rho/3$.  In attempting to identify the reasons why
$\Lambda$CDM produces phases of acceleration and deceleration, one is
reminded of trying to fit a straight line with a low-order polynomial---it
is always possible to make the ends meet, but there will be inevitable
wiggles in between.  It appears that the early deceleration and current
acceleration indicated by $\Lambda$CDM are two of these wiggles, whereas
$R_{\rm h}=ct$ fits the straight line perfectly.

In this work, we have also demonstrated that the class of Type~Ia SNe
continues to be critical to our understanding of how the Universe evolves,
the results reported here affirming the expectation from theory and general
relativity that only a perfect fluid with zero active mass (i.e.,
${\rho+3p=0}$) can be consistent with the use of an FRW metric (Melia
2013b).

\acknowledgments We are very grateful to the anonymous referee for
carefully reading the manuscript and making several important suggestions
that have improved it significantly. We appreciate the use of the public
data from the Supernova Legacy Survey. We also thank Bo~Yu and Fa-Yin Wang
for helpful discussions. XFW~acknowledges the National Basic Research
Program (`973' Program) of China (grants 2014CB845800 and 2013CB834900),
the National Natural Science Foundation of China (grants nos.\ 11322328 and
11373068), the One-Hundred-Talents Program, the Youth Innovation Promotion
Association, and the Strategic Priority Research Program ``The Emergence of
Cosmological Structures" (grant no.\ XDB09000000) of of the Chinese Academy
of Sciences, and the Natural Science Foundation of Jiangsu Province (grant
no.\ BK2012890).  FM~is grateful to Amherst College for its support through
a John Woodruff Simpson Lectureship, and to
Purple Mountain Observatory in Nanjing, China, for its
hospitality while this work was being carried~out.  This work was partially
supported by grant 2012T1J0011 from The Chinese Academy of Sciences
Visiting Professorships for Senior International Scientists, and grant
GDJ20120491013 from the Chinese State Administration of Foreign Experts
Affairs.

\end{document}